# Formation and manipulation of domain walls with 2-nm domain periodicity in BaTiO$_3$ without contact electrodes

*Maya Barzilay and Yachin Ivry*[*]

[1]Department of Materials Science and Engineering, Technion – Israel Institute of Technology, Haifa 3200003, Israel

[2]Solid State Institute, Technion – Israel Institute of Technology, Haifa 3200003, Israel

[*] Correspondence to: ivry@technion.ac.il

**Abstract**

Interfaces at the two-dimensional limit in oxide materials exhibit a rich span of functionality that differs significantly from the bulk behavior. Among such interfaces, domain walls in ferroelectrics draw special attention because they can be moved deterministically with external voltage, while they remain at place after voltage removal, paving the way to novel neuromorphic and low-power data-processing technologies. Ferroic domains arise to release strain, which depends on material thickness, following Kittel's scaling law. Hence, a major hurdle is to reduce the device footprint for a given thickness, *i.e.*, to form and move high-density domain walls. Here, we used transmission electron microscopy to produce domain walls with periodicity as high as 2 nm without the use of contact electrodes, while observing their formation and dynamics *in situ* in BaTiO$_3$. Large-area coverage of the engineered domain walls was demonstrated. The domain-wall density was found to increase with increasing effective stress, until arriving to a saturation value that reflects 150-fold effective stress enhancement. Exceeding this value resulted in strain release by domain-wall rotation. In addition to revealing this multiscale strain-releasing mechanism, we offer a device design that allows controllable switching of domain-walls with 2-nm periodicity, reflecting a potential 144-Tb/inch$^2$ neuromorphic network.

## Introduction

Ferroelectrics are functional materials with high dielectric constant that are found in everyday technologies, ranging from ultrasound imaging to cellular communication, and energy-efficient memory devices.[1,2] The basic functional element in ferroelectrics is polarization domains, *i.e.*, region in which all the crystallographic unit cells are distorted collectively, giving rise to correlated dipole moments. The continuous interest in device miniaturization has lately shifted the focus from domains to the domain walls. Not only are ferroic domain walls typically much smaller than the domains themselves, but also these two-dimensional structures exhibit enhanced functionality, which is different than the bulk, such as magnetism[3] and superconductivity.[4] A special attention is given to conductivity,[5–7] mainly because domain walls are mobile and can be formed by external electric field, allowing low-power non-volatile memory or computing devices. For instance, recent works showed that domain-wall conductivity in $LiNbO_3$ is higher by three to four orders of magnitude than the bulk conductivity,[8,9] while charged domain walls have been observed also for the seminal ferroelectric $BaTiO_3$.[10] Likewise, combined memory-computing devices (*i.e.*, memristors) have been demonstrated by moving domain walls electrically in lead zirconate titanate (PZT).[11]

Competitive low-power data-storage technologies require high density domain walls. Nevertheless, the domain width ($w$), and hence domain-wall density, demonstrate strong dependence on the geometry. The dependence of unperturbed-domain width ($w_0$) on thickness ($d$) in ferroics was formulated by Kittel for ferromagnetism and has later been adopted for ferroelastic[12] and ferroelectric[13] materials:

$$w_0^2 = \frac{4\pi^3}{8.42} \frac{(\gamma d)}{G(S_a - S_c)^2} \qquad (1)$$

where $G$ is shear modulus, $\gamma$ is domain-wall energy and $S_a$, and $S_c$ are the respective spontaneous strain along the respective pseudo-cubic $a$ and $c$ axes of the ferroelectric perovskite. Following these scaling law, domains with 27-nm periodicity have been observed in relaxed $PbTiO_3$ films with minimal film-substrate lattice mismatch.[14] Likewise, 10-nm periodicity has been observed in polycrystalline PZT films.[15] In these materials, the domains organize in a long-range order of periodic stripes with common alignment, *i.e.*, in the form of bundle domains.[16] Bundle domains can be switched as a whole,[17,18] allowing collective switching of several domains or several memory cells, simultaneously.

Recent theoretical predictions[19] suggest that substrate-free $BaTiO_3$ can comprise domains as small as ~2 nm. Nevertheless, realization of these predictions is a challenge not only because of the difficulties associated with manipulating such small domains, but also because observing them pushes the capabilities of contemporary microscopy. The most common method for both manipulating and imaging domains, piezoresponse force microscopy (PFM), relies on a conductive electrode that performs a raster scan.[20,21] Nevertheless, the imaging resolution in PFM is limited to ~1 nm at most,[15] while the resolution for domain writing is worse.

An alternative method to manipulate ferroelectric domains is contact-less, *e.g*., using an electron beam. Low-energy beams have been used to form and observe domains using scanning or transmission electron microscopy (SEM and TEM, respectively).[22–26] Moreover, electron microscopy has been used to image in-situ domain-formation dynamics.[27] Nevertheless, low-energy and low-dose electron beam cannot supply high-resolution imaging, neither high-density domains. Contrariwise, high-resolution imaging, at the atomic scale, requires much higher doses,[28] and high energy,[29,30] *e.g*., by using scanning TEM methods. But these irradiation conditions change the material composition and damage the material.[31] Here, we chose a non-conventional zone axis, at which the domain walls are not edge-on, allowing us to observe domain formation conveniently. We demonstrate the formation and manipulation of correlated domain walls with 2-nm domain periodicity in substrate-free $BaTiO_3$ crystallites.

## **Experimental**

50-nm $BaTiO_3$ crystallites were examined. The particles were suspended in ethanol and sprayed on an amorphous Carbon-Cu grid using high pressure $N_2$. Details regarding sample preparation and complementary characterization can be found elsewhere.[31,32]

The samples were then inserted to a double-corrected Titan Themis $G^2$ 60-300 (FEI / Thermo Fisher) HRTEM with sub-angstrom resolution. All experiments were done using 200 keV acceleration voltage and $I = 1$–1.6 nA monochromated currents for $A = 300$-6500-$nm^2$ beam diameter, for both imaging, and domain manipulation.

Because we wished to observe domain formation at the nanometer scale, we wanted to maximize the contrast at the domain wall with respect to the domains, while yet being able to observe the unit-cell tetragonality in the domains themselves. We thus chose the {011} zone axis, in which 90° *a/c* and 180° *c+/c-* domain walls are tilted and hence have a broad areal footprint in their projection at the observed plane. Figure 1 shows the predicted observation of a 90° domain with [-110]$_a$ zone axis. Here, the domain walls are formed diagonally to the beam (Figure 1A) and appear as dark and bright fringes that indicate the presence of the spatial interference of the domain walls in the lattice. From this orientation, the domain-wall projection forms fringes at an angle of 54.8° with respect to the pseudo-cubic structure (Figure 1B).

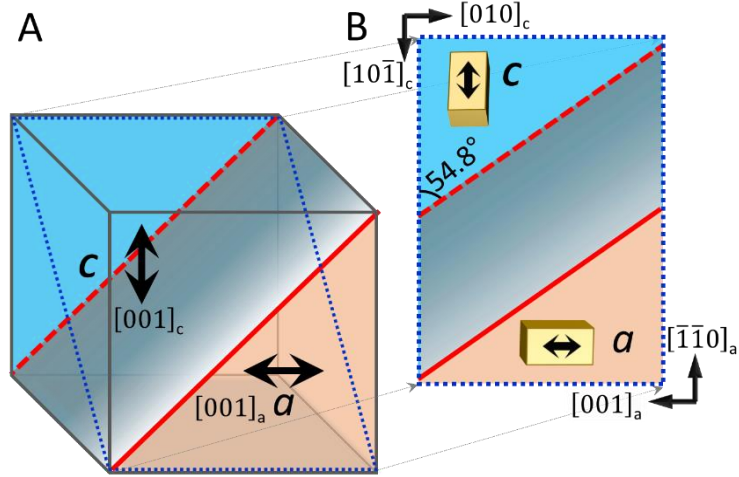

**Figure 1| Crystallographic analysis of 90° domain-wall imaging from a {011} zone axis**. (**A**) Schematics of *a* and *c* unit cells that comprise a 90° domain wall. (**B**) Projection of the structure in (A) on the plane that is perpendicular to the TEM electron beam. The wide area at which the domain wall is expected to form fringes between the pure *a* and pure *c* domains is designated in a white-gray gradient, while the angle between the projected wall and the *a* or *c* domains with respect to the pseudo-cubic structure is also marked (54.8°). Dotted blue frame, as well as continuous, and dashed red lines in (B) and (A) are matching, to help guide the eye.

## Results and Discussion

Following the scaling law of Equation 1, as well as earlier experimental works,[33–35] the anticipated domain periodicity in a 50-nm BaTiO$_3$ crystallite ($\gamma \approx 3 \times 10^{-3}$ J m$^{-2}$) is about 25 nm. For this estimation, we consider the thickness of the crystallite as about half of its diameter. Figure 2A shows a TEM image of a < 50-nm BaTiO$_3$ crystallite, while no 90° domain walls appear from the observed {011} orientation.

To form domain walls, we exposed the crystallites to a constant current that went through the sample. Hence, increasing the applied dose is proportional to the accumulated exposure time. The crystallites were electrically floating and not connected to any electric-circuit ground within the TEM. Figures 2A-C show the formation and evolution of domain walls with 2.5-nm domain periodicity as a function of increasing exposure time (*i.e.*, charging). Similarly, Figure 2C shows the angle between the formed domain walls and the pseudo-cubic [110] directions. This measured angle is 54°, similar to the expected 54.87° calculated angle to be seen from [-110] orientation (Figure 1B) and hence indicating that the bundle domain comprises periodic *a-c* 90° ferroelastic domains. The interference fringes, caused by the formation of domain walls, started to show up after only a few minutes of exposure. The interference stripes are initially formed perpendicular to the particle surface, as known to be energetically preferable in a case of crystallite boundary.[36] Schematics of the contact-less domain formation process under the beam are given in Figures 2D-F.

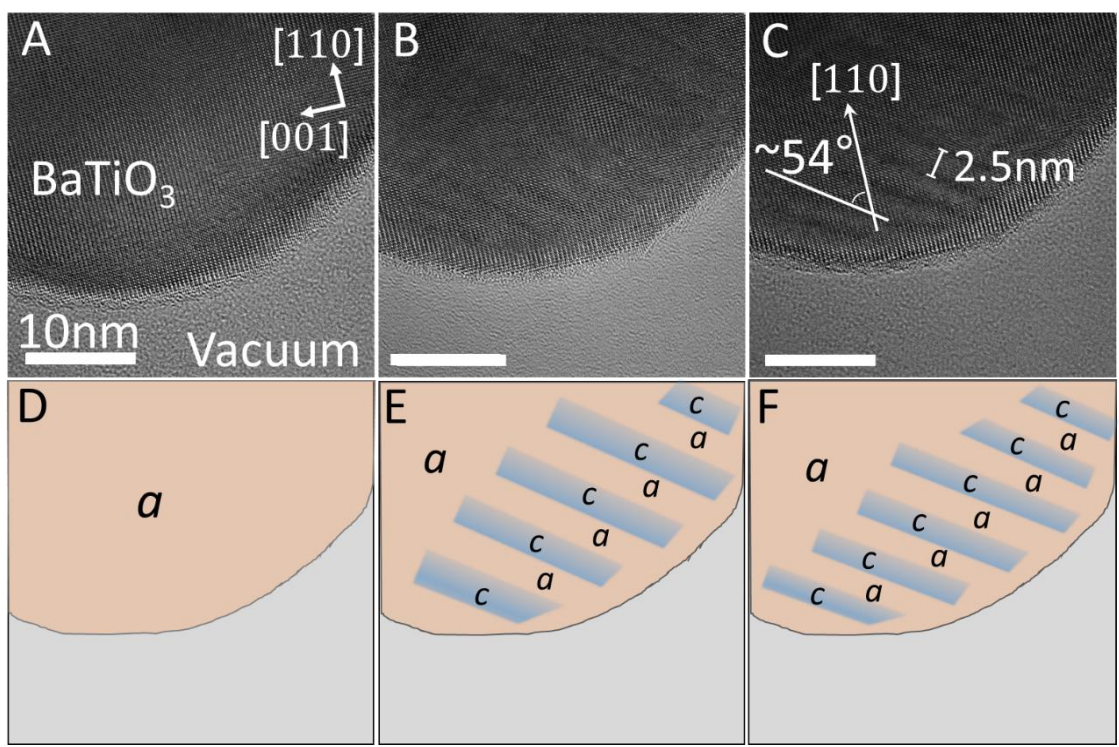

**Figure 2| Domain-wall formation with 2.5-nm spacing in BaTiO₃.** (**A**) TEM image of a < 50-nm BaTiO₃ crystallite shows no domain walls in the native structure. (**B**) Domain-wall growth under low-energy electron exposure for 556 and (**C**) 1145 seconds. (**D-F**) Schematic illustrations of the domain formation process, which correspond the HRTEM images in (A-C). A ~54° angle between the domain walls and the pseudocubic orientation that is designated in (C) and (F) corresponds to the expected angle in Figure 1B.

To verify the crystallographic domain analysis, we demonstrated the disappearance of these domains above the Curie temperature as well as characterized the change in lattice parameter in the artificially formed domains. Figure 3A shows the absence of small domains in the pristine crystallite, followed by periodic domain (Figure 3B) that were formed by exposing the crystallite to the electron beam for 600 seconds. These domains disappeared completely upon heating the sample above the Curie temperature, as seen in Figure 3D.[32]

Figure 3B helped us also verify the formed *a* and *c* domains structure in the crystallite, by direct characterization of the tetragonality in the unit cells. Given the predicted projection of an *a-c* domain wall from the zone axis used in this work (Figure 1B), we expect to measure the following across the interference fringe: a few unit cells of pure *a* domains (*i.e.*, long inter-atomic distance); mixed *a-c* domains due to the tilt of the domain wall with respect to the beam direction; and pure *c* domains (*i.e.*, sort inter-atomic distance). Figure 3C shows the interatomic spacing along the [001] direction in the area designated in yellow. This measured interatomic-distance distribution agrees with the above expectation.

The domain formation is a gradual process. That is, the domain periodicity increases with increasing exposure time, so that the domain width can be determined by stopping the excitation at any given time. Figure 3E shows the domain-width evolution with exposure time for several crystallite excitations, while a constant average dosage of 1.5 nA/nm$^2$ was used. The ferroelastic domain width in the examined crystallites has a saturation minimal size as small as 2 nm (about five unit cells). Irradiating the material when the domains arrive to this saturation value does not change any more the domain periodicity. The existence of such a saturation value indicates that at this value, the domains are in equilibrium. Once formed, the small domains with the 2-nm periodicity remained unchanged even if the electron beam did not irradiate the area for a long period of time (*e.g.*, see gap in data acquisition between 5622 and 7685 sec in Figure 3E). The data suggest that although the change in periodicity is rather gradual, one can distinguish between three main domain-width sizes: large (~8 nm) intermediate (~4 nm) and saturation (2 nm). This implies that the basic mechanism is likely to be related to domain doubling,[37] while domain-wall mobility may play a role in the gradual change in periodicity.

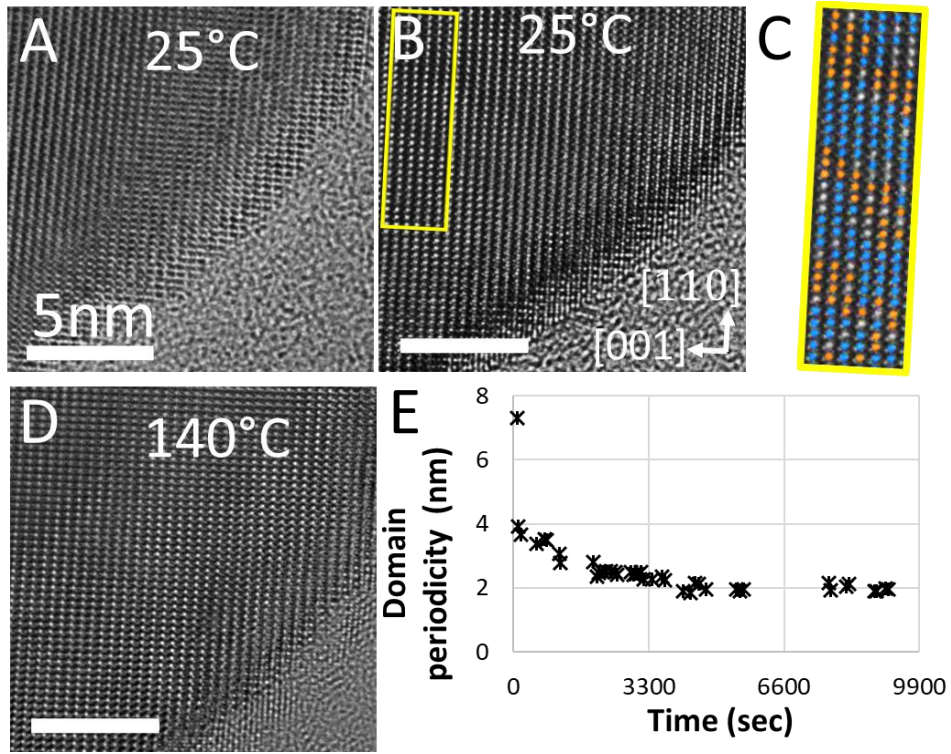

**Figure 3| Temperature dependence of nanometer-scale engineered domains.** (**A**) TEM image at room temperature of a pristine < 50-nm BaTiO$_3$ crystallite, showing no domains. (**B**) The same sample with domain walls that were formed after 600 seconds of exposure. These walls are perpendicular to the edge of the crystallite. (**C**) Closer look at the highlighted area, in which the atomic spacing was measured in each unit cell. Blue and orange superimposed unit-cell coloring correspond to $c$ domains (3.95 $\pm$0.05 Å spacing) and $a$ domains (4.05$\pm$0.05 Å spacing), respectively. Regions with 4.00-4.028 Å spacing due to *e.g.*, *a-c* domain overlap are colored gray. See error estimation in Figure SI1. (**D**) At 140° C, the tetragonal structure becomes cubic and the domains are eliminated. (**E**) Domain periodicity as a function of exposure time for several samples, showing a decrease from 8 nm to a saturation periodicity of 2 nm. Here, time starts running from the formation of the first domains.

The interaction between an electron beam and the ferroelectric crystal can excite the material in various ways. The electron beam that impinges the ferroelectric gives rise to an electric field that switch the domains.[26] The electric field may affect the domain structure also indirectly, *i.e.*, by inducing stress or strain locally, due to the strong electro-mechanical coupling in these piezoelectric and flexoelectric materials.[15,38,39] Because 90° domains arise to release strain they are absent in the case of fully relaxed samples, similarly to the absence of 180° polarization domains when a ferroelectric is short circuited.[40] Hence, formation of 90° domains indicates that the mechanism that induced them involves introduction of stress. The value of this stress is extracted directly from Equation 1 for equilibrium states. That is, the observed 2 nm periodicity corresponds to structures as thin as ~1 nm, much smaller than the 25-nm periodicity that is expected for a 50-nm BaTiO$_3$ crystallite.[19,34] Hence, we can conclude that the reduction in domain periodicity is due to excess shear stress that was induced by the beam, either directly or indirectly due to, *e.g.*, piezoelectricity and local heating. The enhancement in shear stress is proportional to the squared ratio between the anticipated and observed ($w_{\text{measured}}$) periodicities (Equation 1): $G_{\text{eff}}/G = w_0/w_{\text{measured}}$ , *i.e.*, the electron-beam excitation is equivalent to an enhancement of $G_{\text{eff}} = 150G$ with respect, *e.g.* to a crystalline film of the same thickness.

Because ferroelastic (non-180°) domains arise to minimize mechanical energy, the existence of a minimal domain width (2 nm) indicates on a maximal stress value that can be released by this domain-periodicity doubling. Therefore, we expect that for times longer than that requires to arrive the saturation value (~3300 sec, which translates into 5.4x10$^{-6}$ C for the global charge injection in $A$ = 6361 nm$^2$), the ferroelectric material will have a different mechanism to release the excess stress. Such additional strain-releasing mechanism can therefore reduce the global stress, *i.e.*, giving rise to a set of bundle-domains[16] at a scale of ~10 nm, which is greater than the 2-nm *a-c* periodicity (Figure 2E). Using this concept, we wanted to show that the above method of contact-less electron-beam excitation is useful also for forming high-density domains over a much larger scale. Figure 3A shows a crystallite with only a few domain walls (far away from saturation periodicity). Here, the domain walls are perpendicular to the edge of the crystallite as in the above samples (Figure 2). Exciting the crystallite with electron beam for 3000 sec. generated high-density domains that covered nearly the entire material (Figure 3B), demonstrating scalability of the high-density domain formation.

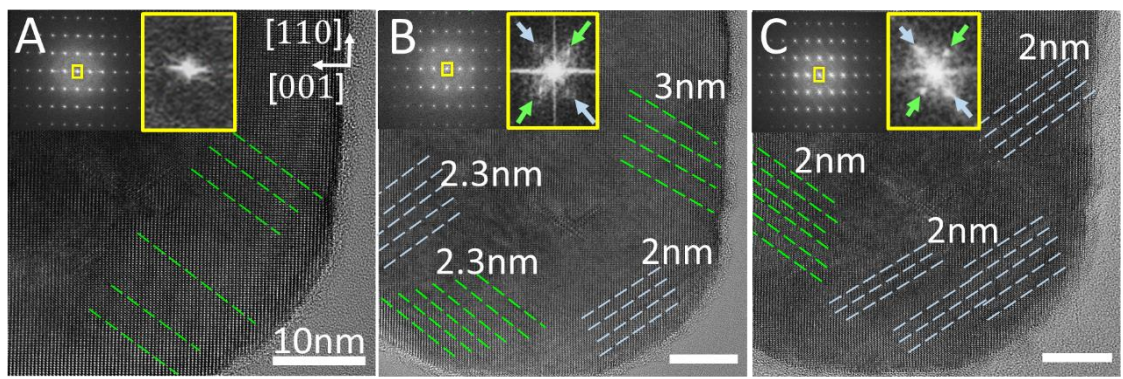

**Figure 4| Large-scale formation and switching of high-density domain-wall structures**. (**A**) Large-scale TEM image of a BaTiO$_3$ crystallite with minimal exposure to the beam. Inset: FFT of the ordered atomic structure, showing a single peak at the Fourier space that corresponds to a specific atomic location in the crystal structure. The perpendicular

orientation of domain walls with respect to the geometrical boundary of the sample is energetically favorable (note that some domain walls were formed during the relatively long beam-alignment duration). (**B**) Increasing the dose (3000 sec exposure) induced high-density periodic domains with long-range coherency (bundle domains) that span the entire crystallite. (**C**) After 8100-sec exposure, the domain periodicity reached its saturation, giving rise to strain release in the form of domain-wall rotation. Here, the domain rotation was induced when the beam was concentrated at areas in which the domain walls are parallel, and not perpendicular, to the edge of the crystallite [*e.g.* bottom-left corner in (B)]. The corresponding FFT data [inset in (B) and (C)] show satellite frequencies around the atomic-structure fingerprint. These satellite frequencies correspond to 2-2.5-nm spacing, at an angle of 54° with respect to the pseudo-cubic structure. Domain widths of specific bundle domains are marked on the micrographs.

Observing the newly formed domains over a large scale allowed us to perform a reliable statistical analysis, such as fast Fourier transform (FFT). The FFT images of Figure 4A (insets) correspond to the spatial frequencies of the atoms in the crystal. However, a close look at the FFT of the crystal with the large-scale domains (yellow frame in Figure 4B) shows satellite peaks around the peaks that correspond to the atomic-scale pseudo-cubic structure. These satellite peaks (0.47 $nm^{-1}$) match the large-scale domain periodicity in the crystallite. That is, the satellite periodicities represent 2.5-nm domain size, while the projection of the domain-wall orientation with respect to the pseudo-cubic orientation is ~55°, in agreement with the prediction (Figure 1B) as well as with Figure 2C.

Utilizing domain walls for devices requires controllable domain-wall mobility. Hence, we wished to demonstrate that the high-density domain-wall structures are switchable. Previous PFM studies show that organized periodic domains can be switched when they are excited near the bundle domain wall or the crystallite boundary.[41] Moreover, in crystallites, the favorable energetic organization of striped 90° domains is having the domain walls perpendicular to the geometrical boundary.[42] Thus, walls that are not perpendicular to the crystallite boundary reflect and unfavorable energy state and hence are expected to be switchable.[42]

Figure 3B shows a $BaTiO_3$ crystallite that comprises bundles of striped 90° domains that were formed under the electron beam. To switch these domains, we focused the beam at regions, in which the domain walls were not perpendicular to the geometrical boundary (*e.g.*, at the bottom-left region of the crystal). Figure 3C shows that the excitation switched the domains successfully. The 90° domain walls became perpendicular to the excited geometrical boundary after the switching.

Bundle-domain rotation occurred after the domains arrived to their saturation size as well as after that bundle domains are formed. Hence, this rotation can be accounted for as a large-scale stress-releasing mechanism that completes the domain doubling. Therefore, the electron-beam irradiation results in a multiscale stress-releasing mechanism, so that small amount of charge injection (between $1.3 \times 10^{-7}$ C and $5.4 \times 10^{-6}$ C for $A = 6361$ nm$^2$ ) gives rise to domain formation. This process continues until the domains reach a saturation width (2 nm in Figure 3E and Figure 4). Increasing the charge injection further ($> 5.4 \times 10^{-6}$ C for $A = 6361$ nm$^2$), induces larger-scale domain switching. Hence, the method demonstrated here is useful for systems that require high-density domain walls, systems that rely on domain-rotation engineering, as well as systems that need both of these mechanisms. Schematic of a proposed design for such a switching device is given in Figure 5.

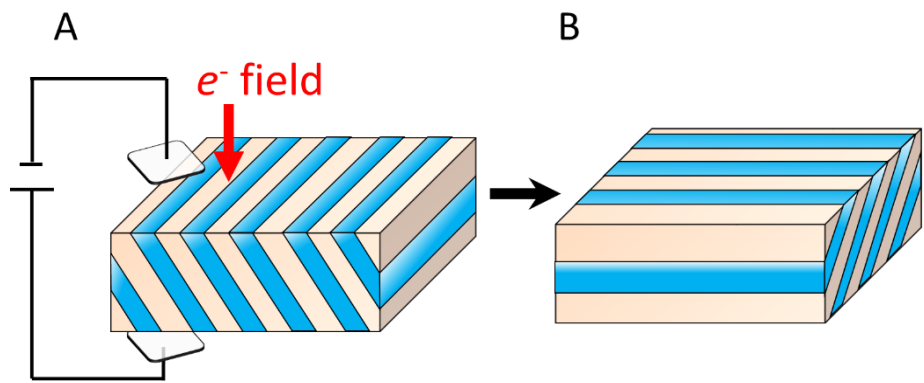

**Figure 5| Geometry-assisted domain switching**. (**A**) Schematic illustration of domain distribution in a structure with finite geometry. In this geometry, the domain walls are perpendicular to some of the edges (favorable energetic state), but are parallel to others. In this geometry, it is possible to excite domains at regions with domain walls parallel to the edge. (**B**) Excitation of domain walls that are not perpendicular to the edge gave rise to domain rotation, so that the domain walls became perpendicular to the edge at the excited area. Further excitation of the same regions does not affect the domain orientation. However, as a result of the rotation, the domain walls became now parallel to a different boundary. Thus, exciting these regions again will switch the domains, giving rise to a reversible-switching process.

## Conclusion

It has been demonstrated that domains in the seminal ferroelectric BaTiO$_3$ can be engineered to a size as small as 2 nm. The domain walls are induced by will, Domain formation can be induced even without the use of contact electrodes. The effective enhancement in shear stress that forms these domains was calculated and found to be of a factor of ~150. Higher hierarchy of domain organization in the form of bundle domains have also been illustrated. Given the usage of domain walls in modern data storage and processing devices, these results are promising for high-density data-processing technologies.

We should note that we could not obtain complementary images of the domains and domain walls from angles perpendicular to the zone axis used in the above images. The reason is that such imaging requires good mechanical stabilization of the crystals. However, upon rotation of the sample, the samples move slightly on the grid so that eucentric rotation is impossible. Therefore, currently, it cannot be determined whether a formed domain wall is charged or uncharged nor can we tell whether the domain wall is accompanied by field closure of vortex-like structures.[43,44] Likewise, it is impossible to conclude what the domain-wall width, which should be narrower than the domain width, though we believe that previous works show domain-wall width of ~1 nm for such domains (see Figure 2D in Reference).[31] Yet, the existence of a large number of domain walls in various orientations strongly suggests that at least some of them are charged and exhibit high conductivity. Moreover, the long-range correlation of domain walls allows collective switching of a large number of domain walls simultaneously. Finally, the general method that is presented here to design devices that make use of their geometry for obtaining collective domain switching is applicable either with or without dedicated contact electrodes.

## **Acknowledgments**

We acknowledge support from the Zuckerman STEM Leadership Program, the Russel Barry Nanoscience Institute, and the Israel Science Foundation (ISF) grant #1602/17. Likewise, we would like to thank Dr. Yaron Kauffmann, Mr. Michael Kalina and Mr. Elad Barzilay for technical support.

## **References**

1    Y. Ivry, N. Wang, C. Durkan, *Appl. Phys. Lett.* 2014, **104**, 133505.

2    E. Y. Tsymbal, A. Gruverman, *Nat. Mater.* 2013, **12**, 602.

3    X. K. Wei, A. K. Tagantsev, A. Kvasov, K. Roleder, C. L. Jia, N. Setter, *Nat. Commun.* 2014, **5**, 3031.

4    A. Aird, E. K. H. Salje, *J. Phys. Condens. Matter* 1998, **10**, L377.

5    J. Seidel, P. Maksymovych, Y. Batra, A. Katan, S. Y. Yang, Q. He, A. P. Baddorf, S. V. Kalinin, C. H. Yang, J. C. Yang, Y. H. Chu, E. K. H. Salje, H. Wormeester, M. Salmeron, R. Ramesh, *Phys. Rev. Lett.* 2010, **105**, 197603.

6    J. Seidel, L. W. Martin, Q. He, Q. Zhan, Y.H. Chu, A. Rother, M. E. Hawkridge, P.

Maksymovych, P. Yu, M. Gajek, *Nat. Mater.* 2009, **8**, 229.

7   J. Seidel, P. Maksymovych, Y. Batra, A. Katan, S.Y. Yang, Q. He, A. P. Baddorf, S. V. Kalinin, C.H. Yang, J.C. Yang, Y.H. Chu, E. K. H. Salje, H. Wormeester, M. Salmeron, R. Ramesh, *Phys. Rev. Lett.* 2010, **105**, 197603.

8   C. Godau, T. Kämpfe, A. Thiessen, L. M. Eng, A. Haußmann, *ACS Nano* 2017, **11**, 4816.

9   H. Lu, Y. Tan, J. P. V. McConville, Z. Ahmadi, B. Wang, M. Conroy, K. Moore, U. Bangert, J. E. Shield, L. Chen, J. M. Gregg, A. Gruverman, *Adv. Mater.* 2019, **31**, 1902890.

10  T. Sluka, A. K. Tagantsev, P. Bednyakov, N. Setter, *Nat. Commun.* 2013, **4**, 1808.

11  P. Sharma, Q. Zhang, D. Sando, C. H. Lei, Y. Liu, J. Li, V. Nagarajan, J. Seidel, *Sci. Adv.* 2017, **3**, e1700512.

12  A. L. Roitburd, *Phys. status solidi.* 1976, **37**, 329.

13  A. Schilling, T. B. Adams, R. M. Bowman, J. M. Gregg, G. Catalan, J. F. Scott, *Phys. Rev. B.* 2006, **74**, 024115.

14  A. H. G. Vlooswijk, B. Noheda, G. Catalan, A. Janssens, B. Barcones, G. Rijnders, D. H. A. Blank, S. Venkatesan, B. Kooi, J. T. M. De Hosson, *Appl. Phys. Lett.* 2007, **91**, 112901.

15  Y. Ivry, D. P. Chu, J. F. Scott, C. Durkan, *Phys. Rev. Lett.* 2010, **104**, 207602.

16  Y. Ivry, D. P. Chu, C. Durkan, *Nanotechnology* 2010, **21**, 065702.

17  J. F. Scott, A. Hershkovitz, Y. Ivry, H. Lu, A. Gruverman, J. M. Gregg, *Appl. Phys. Rev.* 2017, **4**, 041104.

18  Y. Ivry, J. F. Scott, E. K. H. Salje, C. Durkan, *Phys. Rev. B.* 2012, **86**, 205428.

19  J. Mangeri, Y. Espinal, A. Jokisaari, S. Pamir Alpay, S. Nakhmanson, O. Heinonen, *Nanoscale* 2017, **9**, 1616.

20  A. Gruverman, M. Alexe, D. Meier. *Nat. Commun.* 2019, **10**, 1661.

21  S. V. Kalinin, A. N. Morozovska, L. Q. Chen, B. J. Rodriguez, *Reports Prog. Phys.* 2010, **73**, 056502.

22  A. C. G. Nutt, V. Gopalan, M. C. Gupta, *Appl. Phys. Lett.* 1992, **60**, 2828.

23  M. C. Gupta, W. P. Risk, A. C. G. Nutt, S. D. Lau, *Appl. Phys. Lett.* 1993, **63**, 1167.

24  J. E. Rault, T. O. Menteş,  a Locatelli, N. Barrett, *Sci. Rep.* 2014, **4**, 6792.

25  D. B. Li, D. R. Strachan, J. H. Ferris, D. A. Bonnell, *J. Mater. Res.* 2006, **21**, 935.

26  J. L. Hart, S. Liu, A. C. Lang, A. Hubert, A. Zukauskas, C. Canalias, R. Beanland, A. M. Rappe, M. Arredondo, M. L. Taheri, *Phys. Rev. B* 2016, **94**, 174104.

27  C. T. Nelson, P. Gao, J. R. Jokisaari, C. Heikes, C. Adamo, A. Melville, S.H. Baek, C. M. Folkman, B. Winchester, Y. Gu, Y. Liu, K. Zhang, E. Wang, J. Li, L.Q. Chen, C.B. Eom, D. G. Schlom, X. Pan, *Science.* 2011, **334**, 968-971.

28  C.L. Jia, V. Nagarajan, J.Q. He, L. Houben, T. Zhao, R. Ramesh, K. Urban, R. Waser,

*Nat. Mater.* 2006, **6**, 64.

29  R. Ahluwalia, N. Ng, A. Schilling, R. G. P. McQuaid, D. M. Evans, J. M. Gregg, D. J. Srolovitz, J. F. Scott, *Phys. Rev. Lett.* 2013, **111**, 165702.

30  T. Matsumoto, M. Okamoto, *J. Appl. Phys.* 2011, **109**, 014104.

31  M. Barzilay, T. Qiu, A. M. Rappe, Y. Ivry, *Adv. Funct. Mater.* 2019**,** 1902549.

32  M. Barzilay, H. Elangovan, Y. Ivry, *ACS Appl. Electron. Mater.* 2019, **1**, 2431.

33  A. Schilling, T. Adams, R. Bowman, J. Gregg, G. Catalan, J. Scott, *Phys. Rev. B.* 2006, **74**, 024115.

34  G. Catalan, A. Schilling, J. F. Scott, J. M. Gregg, *J. Phys. Condens. Matter* 2007, **19***, 022201.

35  A. Hershkovitz, F. Johann, M. Barzilay, A. H. Avidor, Y. Ivry, *Acta Mater.* 2020**, 187**, 186-190.

36  Y. Ivry, J. F. Scott, E. K. H. Salje, C. Durkan, *Phys. Rev. B.* 2012, **86**, 205428.

37  A. S. Everhardt, S. Damerio, J. A. Zorn, S. Zhou, N. Domingo, G. Catalan, E. K. H. Salje, L. Q. Chen, B. Noheda, *Phys. Rev. Lett.* 2019, **123**, 087603**.**

38  R. K. Vasudevan, Y.C. Chen, H.-H. Tai, N. Balke, P. Wu, S. Bhattacharya, L. Q. Chen, Y.H. Chu, I.N. Lin, S. V Kalinin, V. Nagarajan, *ACS Nano* 2011, **5**, 879.

39  E. Snoeck, B. Noheda, D. H. A. Blank, A. Lubk, A. Janssens, G. Rijnders, C. Magen, G. Rispens, G. Catalan, A. H. G. Vlooswijk, *Nat. Mater.* 2011, **10**, 963.

40  A. L. Roitburd, *Phys. Status Solidi.* 1976, **37**, 329.

41  C. Durkan, A. Hershkovitz, D. Chu, J. F. Scott, Y. Ivry, *arXiv,* 2016, **1608.03890**.

42  Y. Ivry, N. Wang, D. Chu, C. Durkan, *Phys. Rev. B.* 2010, **81**, 174118.

43  Y. Ivry, D. P. Chu, J. F. Scott, C. Durkan, *Phys. Rev. Lett.* 2010, **104**, 207602.

44  A. K. Yadav, C. T. Nelson, S. L. Hsu, Z. Hong, J. D. Clarkson, C. M. Schlepuëtz, A. R. Damodaran, P. Shafer, E. Arenholz, L. R. Dedon, D. Chen, A. Vishwanath, A. M. Minor, L. Q. Chen, J. F. Scott, L. W. Martin, R. Ramesh, *Nature* 2016, **530**, 198.

**Supplementary information**

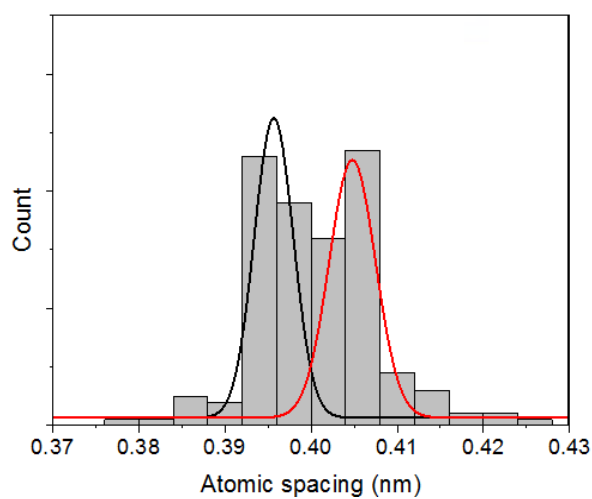

**Figure SI1| Unit-cell spacing distribution and error estimation.** Histogram of the unit-cell spacing distribution extracted from Figure 3C. Black and red curves are the best fit for Gaussian functions that are centered respectively at 3.95 Å, and 4.05 Å with 0.025 Å standard deviation. These data define the atomic spacing at *a* and *c* domains, as well as the standard error.

**Video Link**

**Video SI1| High-density domain wall formation, evolution and rotation in BaTiO₃.** High-density domain walls are formed under an electron beam. Domain density increases with increasing exposure time, until reaching a saturation value of 2 nm. Beyond this value, increasing exposure time rotates the domains. Pre-designed domain switching is obtained by using the geometry of the crystallite and exposing intentionally regions, in which the striped domains are not perpendicular to the geometric boundary.